\documentclass[11pt,a4paper]{article}

\usepackage{graphicx}
\usepackage{amsmath}
\usepackage{natbib}
\usepackage{geometry}
\title{Optical Spectroscopy of the IR Source CPM 19 and Surrounding Objects}

\author{
T.~Yu.~Magakian$^{1}$,
T.~A.~Movsessian$^{1}$,
A.~V.~Moiseev$^{2}$,
T.~S.~Molyarova$^{3}$,
R.~I.~Uklein$^{2}$
}

\date{}

\begin{document}

\maketitle

\begin{center}
$^{1}$Byurakan Astrophysical Observatory, Armenia\\
$^{2}$Special Astrophysical Observatory, Russia\\
$^{3}$School of Physics and Astronomy, University of Leeds, UK
\end{center}

\begin{abstract}
Optical spectra of the well-known infrared source CPM~19, which exhibited a strong decline in brightness during the period from 1984--1987 to 2000--2005, have been obtained for the first time. A strong and broad H$\alpha$ emission line has been detected, along with the possible presence of [S\,II] emission. No traces of an absorption spectrum are observed. It is suggested that the optical component of CPM~19 is in the pre-main-sequence stage. Various explanations of the observed properties are considered; a plausible scenario is that CPM 19 may belong to the class of UX~Ori-type stars with an unusually long eclipse duration, similar to that observed in V1184 Tau. Spectra of other nebulous objects in the vicinity of CPM~19, including the HH objects HH~940 and HH~941, have also been obtained and discussed.


\end{abstract}

\section{Introduction}
The well-known strong infrared source CPM~19 (IRAS~05373+2349, 2MASS~J05402422+2350546) is located in the region of the galactic anticenter. It was identified (though incorrectly) with the very faint red star [1]. The history of its early studies, mainly in infrared and radio range, can be found in [2]. In particular, its association with an H$_2$O maser and molecular hydrogen outflows was established. Actually, these observations confirmed the existence of small star-forming region; its presence was suspected even before, when a possible Herbig--Haro object GGD~4 (or GM~2-4) was detected in this area [3,4]. Further studies have shown that CPM~19 source is located at the center of a relatively compact infrared cluster, indeed representing a star-forming region (see [5] and references therein). A more recent review of these studies is presented in [6]. 

During a search for Herbig-Haro objects with the 2.6-m telescope of the Byurakan Astrophysical Observatory (BAO), CPM~19 was detected as a star in the optical range, despite being classified as a Class~I object and thus expected to be optically invisible [2]. Moreover, this star (Gaia~DR3~3404437884812331904) demonstrated strong variability both in the optical and in the infrared bands. Further progress in the photometric and spectroscopic studies of this star over many years was hampered by its faintness (18--20 mag in the \textit{r} and \textit{i} bands).

In [6], an attempt was made to compile all available optical and infrared photometry, and on the basis of these data a light curve of 2MASS~J05402422+2350546 was constructed in the \textit{J} and \textit{i} bands. The resulting light curve turned out to be very similar to those of eclipsing variable stars.  The authors suggest that an eclipse of the visible star by a dust cloud surrounding the protostar CPM 19 occurred between 1984--1987 and 2000--2005. Since 2005 the brightness of the star has remained constant.  Authors estimate the spectral type of the visible star to be a B1--B8 main-sequence star; however, whether the two objects constitute a physically bound binary system remains unclear. Thus, determining the true spectral type and other properties of the visible star associated with CPM~19 becomes a timely and scientifically interesting task. In this study, we present new photometric and spectroscopic observations of the CPM~19 region using BAO 2.6-m and Special Astrophysical Observatory 6-m BTA telescopes. Besides of CPM~19 itself, we analyze the spectra of the several nebulous objects in the area and discuss implications for their nature.

\section{Observations}
Several spectra of CPM~19 were obtained on November~6,~2016 with the 2.6-m telescope of the Byurakan Astrophysical Observatory using the SCORPIO spectral camera and an EEV 42-40 2048 $\times$ 2048 CCD, with a total exposure time of 3600 s. Additional spectral observations with a higher signal-to-noise ratio were carried out on December~17,~2022 and March~15,~2023 with the 6-m BTA telescope equipped with the multimode focal reducer SCORPIO-2 [7] (see Fig.\ref{field}), using the EEV 261-84 CCD detector.

The VPHG1200@540 grism provided a spectral range of 3650-7200~\AA\ at a resolution of about 5.5~\AA. The slit width was 1\ensuremath{^{\prime\prime}}, and the total exposure time was 3600~s. The slit length was approximately 6.3\ensuremath{^\prime} with a scale of 0.4\ensuremath{^{\prime\prime}} per pixel. The total exposure time was also 3600 s, with seeing of about 1.4\ensuremath{^{\prime\prime}}.

In addition, during 2009--2010 a series of direct images in the $R_c$ and $I_c$ bands was obtained with the 2.6-m BAO telescope. Furthermore, on November 19, 2025, we obtained several direct images with the 1-m Schmidt telescope of the BAO in three broadband filters (g\ensuremath{^\prime}, r\ensuremath{^\prime}, and i\ensuremath{^\prime}), analogous to those of the Sloan Digital Sky Survey, with a total exposure time of 1200 s in each filter.

\section{Results}

\subsection{Photometry of CPM~19}

We estimated the brightness of CPM~19 in the \textit{r} and \textit{i} bands using photometric data of several nearby stars from the Pan-STARRS catalog to calibrate the images. Unfortunately, the quality of the direct images obtained in 2009--2010 is not sufficient to derive reliable photometry; however, rough estimates indicate that during this period the stellar brightness was approximately 19.2 mag in \textit{r} and 17--18 mag in \textit{i}. These values are fully consistent with the results reported in [2].

The observations obtained in 2025 also confirm that the brightness of the star has remained approximately constant over the past decade, with measured magnitudes of r\ensuremath{^\prime} = 19.1 and i\ensuremath{^\prime} = 17.2.

\begin{figure}
  \centering
  \includegraphics[width=200pt]{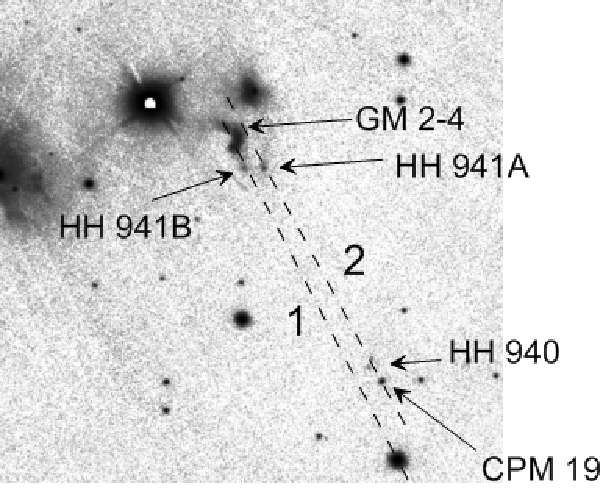}
   \caption{Pan-STARRS \textit{r}-band image of the studied region, showing the positions of the spectrograph slits and the observed objects. Slit 1 corresponds to the observations of December 17, 2022, and slit 2 to those of March 15, 2023.}
   \label{field}
\end{figure}

\subsection{Spectroscopy of CPM~19}

The spectrum of the optical source obtained in 2016 is characterized by a very weak red continuum with a strong superimposed H$\alpha$ emission line. The line has a full width at half maximum (FWHM) of 5.1 \AA (which significantly exceeds the instrumental profile of 2.5 \AA) and exhibits a single-peaked profile without any discernible substructure. The equivalent width is on the order of 60--70 \AA  (this value is uncertain because of the weakness of the continuum), and the radial velocity is $-$11 km s$^{-1}$ (here and throughout the paper, all radial velocities are given in the heliocentric frame).

The spectrum of CPM~19 obtained in 2023 with the 6-m telescope has a higher signal-to-noise ratio, but overall it is virtually indistinguishable from the earlier one. Against the weak continuum, a very strong H$\alpha$  emission line is again observed as a single peak, with an FWHM of 7.8 \AA\  and a radial velocity of +3 km  s$^{-1}$. Its equivalent width is 63 \AA.
Thus, the difference in radial velocities and other parameters between the two spectra is within the limits of observational errors.

\begin{figure}
  \centering
  \includegraphics[width=300pt]{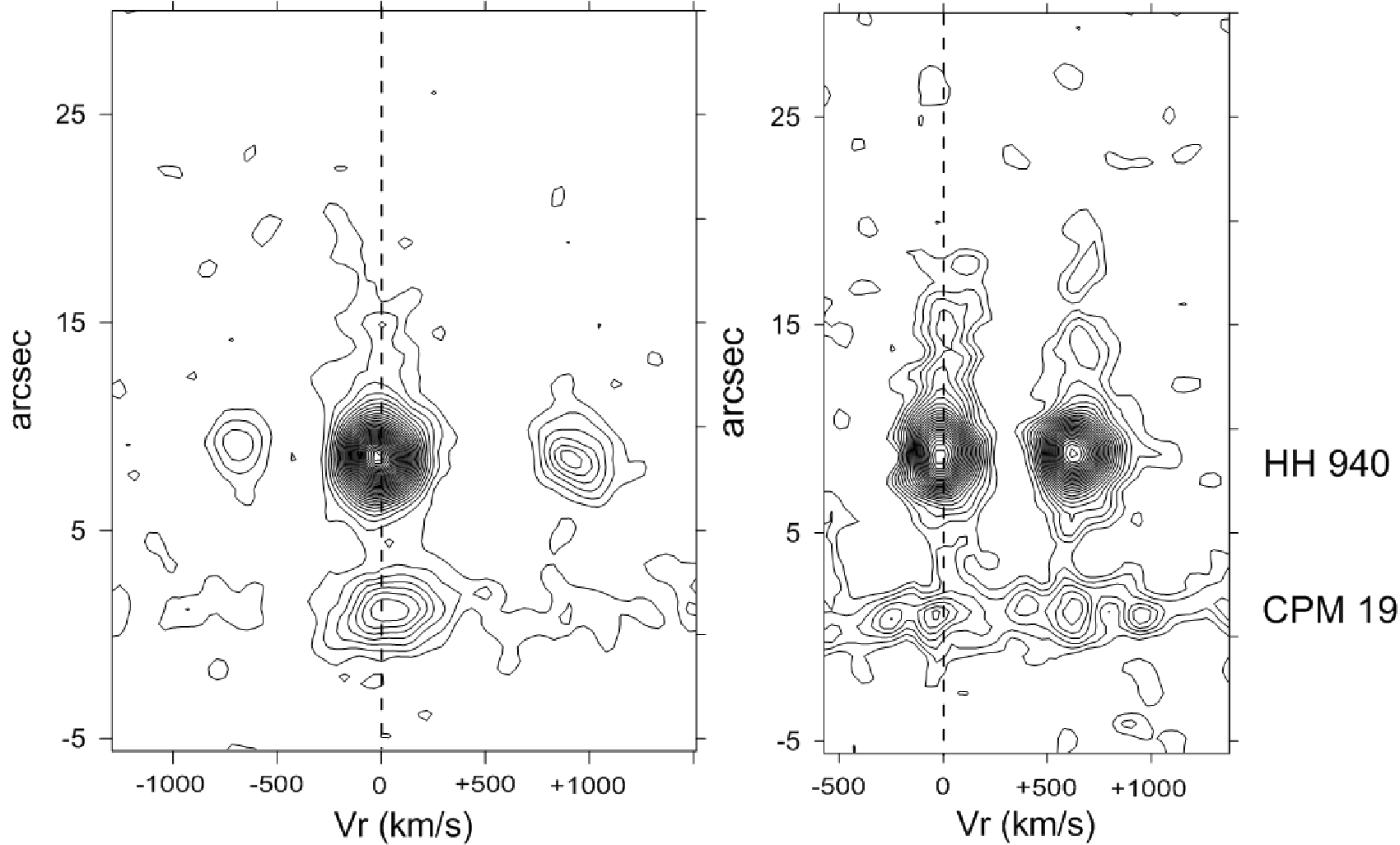}
   \caption{Position-velocity (PV) diagram of CPM~19 and HH~940 in the regions of the H$\alpha$ and [N\,II] lines (left) and the [S\,II] $\lambda$6717 \AA\ and $\lambda$6730 \AA\ lines (right). The near-complete absence of these emission features in the region between CPM~19 and HH~940 is clearly evident, especially in the H$\alpha$  line.}
   \label{CPM}
\end{figure}

It should also be noted that in both spectra -- those obtained in 2016 and in 2023 -- unexpectedly for us the very faint emission from the red [S\,II] doublet was discernible. Moreover, from the 2023 spectrum it seems that these lines may be split into two or three components (see Fig.\ref{CPM}, right). Initially, we suspected that their presence might be due to contamination from the nearby Herbig--Haro object HH~940, located at a distance of less than 10\ensuremath{^{\prime\prime}} from CPM~19; however, a more detailed analysis leads us to reject this assumption (see the section on HH~940 below).

Previously measured spectra from the literature and SED analysis indicate that the CPM~19 system can include a B-type star. SED modelling  performed by [5] (CPM~19 designated as IRS~4) reveals two possible scenarios for the system. The model, which excludes wide-aperture far-infrared and radio observations is consistent with a B-type star with an extinction of about 23
magnitudes. With the FIR data, the object appears significantly less luminous and cooler.
Overall, there may be two objects: a visible source projected against the background of a deeply
embedded core of the infrared cluster [5]. At the same time, the SED analysis performed in [6] yields a single
object of B type with  $A_{\rm V} = 19$ (see their Table~2 for Mol12A). Our 2016 and 2023 optical
spectra, which show the existence of the wide and rather strong   H$\alpha$ emission, are also consistent with the presence of a B-type star.

\section{Spectra of Other Emission Objects}

In the course of our study of CPM~19 in 2022 and 2023, spectra of several other objects in the region were also obtained (Fig.\ref{field}). All of them had been previously identified in [4], and we adopt the designations used in that paper.

\subsection{HH~940}

This object, located close to CPM~19 and consisting of two condensations with a total extent of about 20\ensuremath{^{\prime\prime}}, is described in detail in [2]. Apparently, during our observations condensation B fell within the slit. Its spectrum is entirely typical of low-excitation Herbig--Haro objects and contains emission lines of H$\alpha$, H$\beta$, [S\,II], [N\,II], and [O\,I]. No additional components are detected in these lines, and the radial velocity of condensation B, determined from eight emission lines, is $-$31$\pm$17 km s$^{-1}$. Weak traces of [S\,II] and H$\alpha$  emission are also visible closer to the northwestern edge of HH~940, so that the total length of the object along the slit is about 10\ensuremath{^{\prime\prime}}.

The proximity of HH~940 to CPM~19 suggests that this star might be the driving source of the object. However, no reliable signatures of H$\alpha$ emission are detected in the 6--7\ensuremath{^{\prime\prime}} wide region separating the two objects (Fig.\ref{CPM}). It seems that the presence of a directed HH jet, which connects HH~940 and CPM~19, can apparently be ruled out, which, of course, does not exclude their possible connection. 

\subsection{GM~2-4}

This very red arc-shaped reflection nebula was described in [3] as a possible Herbig--Haro object (see also [4]). It was later noted in [8] that another small reflection nebula is located nearby, designated by those authors as GGD~4a, while GM~2-4 was renamed GGD~4b. In [2], this nebula is described as purely reflection one, with the illuminating star being almost invisible in the optical range; in the near-infrared, however, it is clearly detected as the source 2MASS~05402868+2352304. Even in the Pan-STARRS survey images, which have a significantly deeper limiting magnitude, this star can be discerned only in the \textit{i}, \textit{z}, and \textit{y} filters. Nevertheless, 2MASS~05402868+2352304 exhibits strong H$\alpha$ emission, clearly visible in the reflected spectrum of the nebula: in [2] it was designated as a nebulous emission-line star No. 3. Two HH condensations, HH~941A and HH~941B, are located immediately to the south of GM~2-4. Large-scale images of all these objects are also presented in [5], where the star 2MASS~05402868+2352304 is designated as IRS~7 (their Figs. 2 and 6). During the long-slit observations with the 6-m telescope on December 17, 2022 and March 15, 2023, we obtained spectra of these condensations and of the GM~2-4 nebula.

\begin{figure}
  \centering
  \includegraphics[width=250pt]{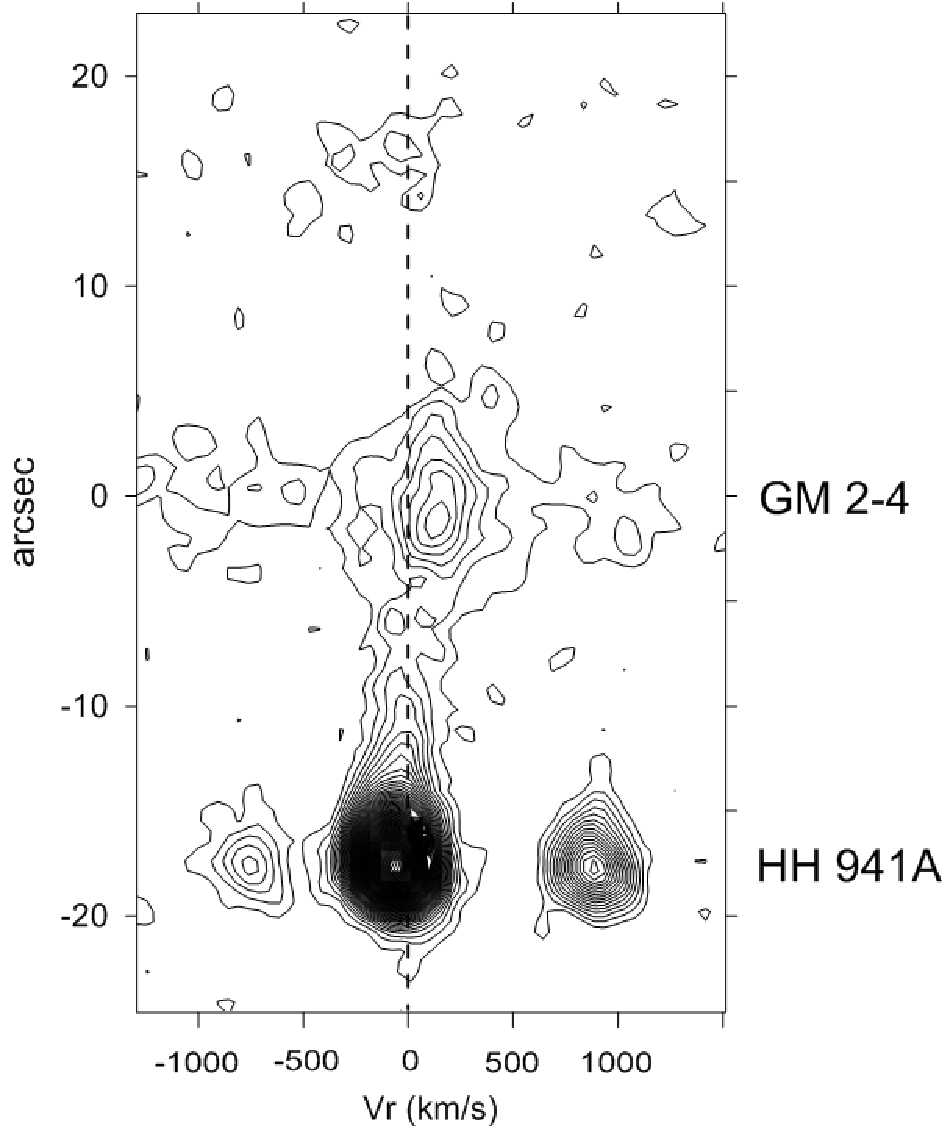}
   \caption{Position-velocity (PV) diagram of the spectra of GM~2-4 (center) and HH~941A (bottom) in the region of the H$\alpha$ and [N\,II] lines. The high positive radial velocity of the H$\alpha$ emission in the spectrum of the nebula is clearly seen, as well as the nearly constant velocity of this emission along the direction from HH~941A toward GM~2-4.}
   \label{GM2-4}
\end{figure}

In our long-slit spectra, GM~2-4 exhibits a well-defined red continuum with FWHM of approximately 6.5\ensuremath{^{\prime\prime}} across dispersion, i.e., clearly broader than the spectrum of a single star. Apparently, we are observing the portion of the nebula illuminated by the star 2MASS~05402868+2352304. Superimposed on this continuum is a strong H$\alpha$  emission line in the form of a single sharp peak, with an equivalent width of 31--32 \AA\  and an unusually high positive radial velocity of +110 $\pm$ 5 km s$^{-1}$(Fig.\ref{GM2-4}). No other features are detected in the H$\alpha$  profile or in the continuum.

\begin{figure}
  \centering
  \includegraphics[width=250pt]{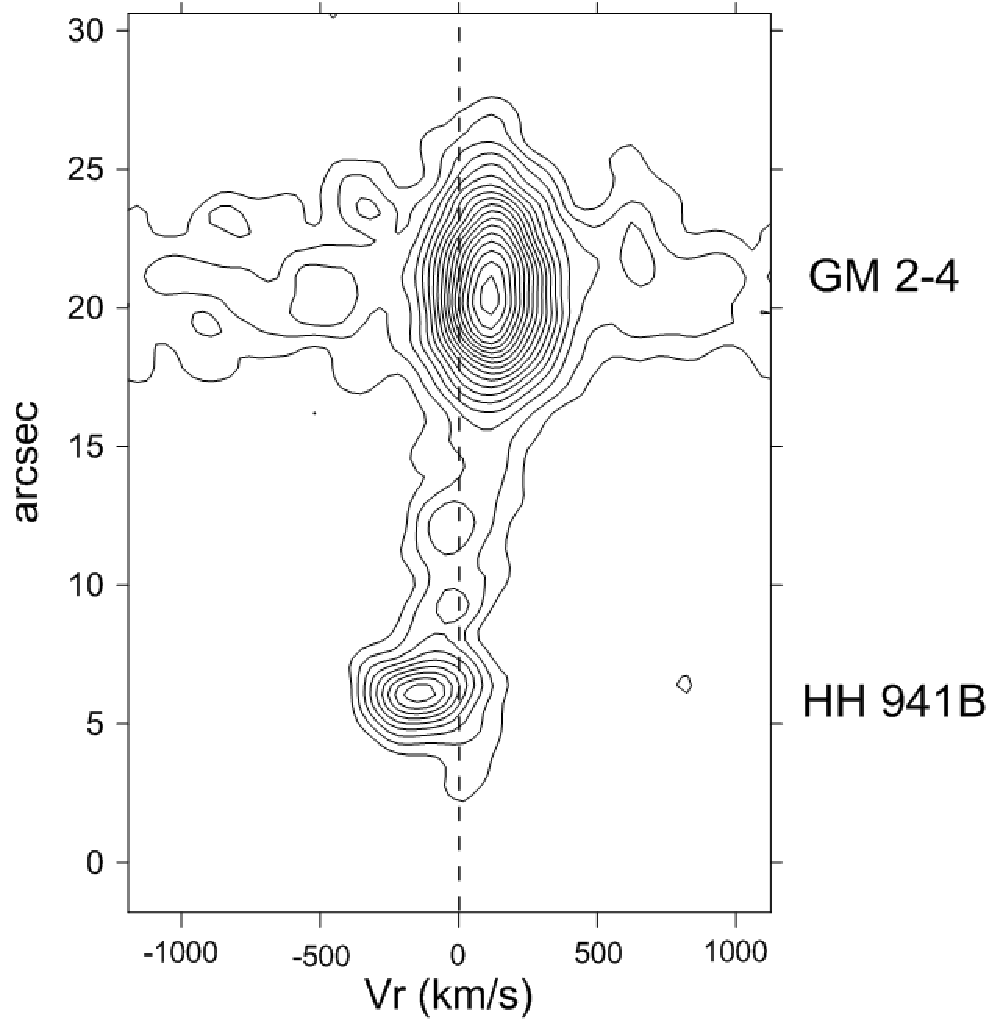}
   \caption{Position-velocity (PV) diagram of the spectra of GM~2-4 (top) and HH~941B (bottom) in the region of the H$\alpha$ line. The high positive radial velocity of the H$\alpha$ emission in the spectrum of the nebula is clearly visible, as well as the rapid increase in the absolute velocity of this emission toward HH~941B (cf. Fig.\ref{GM2-4}).}
   \label{HH941B}
\end{figure}

\subsection{HH~941A and HH~941B}

As noted above, the HH 941 object is located about 16\ensuremath{^{\prime\prime}}  southwest of GM~2-4 and consists of two fairly well-defined knots, A and B, separated by approximately 7\ensuremath{^{\prime\prime}}. We succeeded in obtaining spectra of both knots, which turned out to be markedly different.

HH~941A (the western knot) exhibits a classical Herbig--Haro spectrum, with emission-line intensity ratios virtually identical to those of HH~940. Its radial velocity is $-91 \pm$15 km s$^{-1}$, and no additional structures are observed in the emission-line profiles. However, in the H$\alpha$ and [S\,II] lines faint emission extends from the main condensation almost all the way to the reflection nebula GM~2-4 (Fig.\ref{GM2-4}).

The eastern knot, HH~941B, shows only rather weak H$\alpha$ emission in its spectrum, with a radial velocity of $-$127 km s$^{-1}$. The most intriguing feature of the spectrum, however, is an emission filament connecting HH~941B with the GM~2-4 nebula (presumably -- with the star embedded within it), which is detected spectroscopically in the H$\alpha$ and [S\,II] lines. The [S\,II] lines are very faint but are unambiguously present. In large-scale images of this object (see [5], Fig. 6), a protrusion of condensation HH~941B extending toward GM~2-4 is clearly visible in the [S\,II] lines, and our spectra therefore confirm its emission nature.
The radial velocity of this connecting feature increases in absolute sense smoothly from +61 km s$^{-1}$ near the nebula to $-$50 km s$^{-1}$ at its junction with the main HH object (Fig.\ref{HH941B}). We note that the [S\,II] lines are completely absent in the spectra of the nebula itself and of the HH~941B condensation.

\section{Discussion and conclusion}

The interpretation of the CPM~19 source is complicated by the question of whether it represents a single star or a system of two stars, one of which is presumed to be a Class I object. Both scenarios were proposed in [5] to explain the modeling of the spectral energy distribution (SED). A single star model is consistent with a B-type star
with high visual extinction of $A_{\rm V}=19-23$ [5,6]. In [6], the visual variability of CPM~19 was interpreted as an eclipse of an early-type star by a dense dusty disk surrounding a Class I type object.

The described scenario involving an eclipse by the disk or envelope of a neighboring star appears to be plausible. Let us assume that the objects are not gravitationally bound. If we adopt their relative velocity (v) to be 30 km  s$^{-1}$ (the plausible estimate of the local relative velocity between stars)  and the duration of the eclipse (t) to be 15 years, then the size of the eclipsing object must be $t \times v\approx $100 AU. A hundred astronomical units is a typical size of a protoplanetary disk [9], so in principle this scenario is feasible.

This approach, however, is clearly unlikely, as the eclipsing object would have had to pass very precisely between the observer and the eclipsed star. Nevertheless, in forming clusters of young stars, the typical separations between objects are more often thousands to tens of thousands of AU, rather than the $\sim10^5$ AU characteristic of field stars, which somewhat relaxes this constraint. An alternative possibility is to assume that the objects are components of a binary system. In this case, their relative velocities would be much lower, for example $\sim 3$ km s$^{-1}$ and the required disk size would be correspondingly smaller ($\sim$ 10 AU). Such a size is somewhat smaller than typical disk radii, but it is quite typical for binary systems [10].
Another
possibility is the eclipse by a disc wind, which can serve as a source of optical extinction [11].
In eclipsing binaries, the consideration of the disc wind allows for a wider range of system
inclination angles and relaxes the limits on the relative velocities and disk sizes [12].

On the other hand, the optical+IR source may also be a single star. Taking into account our spectroscopic data -- most notably the presence of fairly strong H$\alpha$ emission and possible signatures of directed outflow, which suggest the youth of the optical star, and without abandoning the eclipse hypothesis, other, somewhat more exotic scenarios can be considered. For example, an eclipse by the star's own protoplanetary disk, as observed in UX Orioni-type objects, i.e., young variable stars predominantly of the Herbig Ae/Be type [13], which is consistent with the
SED fitting [5,6]. The typical duration of their eclipses is much shorter, on the order of weeks; however, in the case of V1184~Tau, an eclipse lasting about 10 years was observed [14], which is already close to the present case. Among the possible explanations discussed in the literature are changes in the accretion rate in a binary system, leading to disk inflation and, consequently, to the obscuration of the star by its dusty component [15]. A more specific scenario proposed
by [16] includes an obscuraion by a dusty disc wind. It was used to explained prolonged minima
of UX Orioni-type objects V1184~Tau and RW~Aur and is indirectly supported by the shallower
depth of the CPM~19 eclipse with increasing wavelengths [6]. Precession of a warped protoplanetary disk with a period of several decades, for example due to the presence of a formed planet [17], may also produce episodic eclipses. For a young star undergoing active planet formation and growth, the presence of a cloud of gas and dust produced by a giant protoplanet collision is also possible, although such events are expected to be shorter-lived, typically up to about a year [18].

The nature of the other IR source, 2MASS~05402868+2352304 -- the star embedded in the GM~2-4 nebula -- is more understandable. According to the SED modeling presented in [5], it is a relatively evolved Class II YSO with a mass slightly exceeding that of the Sun and an active accretion disk. Judging from the reflected spectrum, which shows no absorption features and exhibits strong H$\alpha$ emission, our observations support this conclusion. Somewhat unexpected is the rather large positive radial velocity of the H$\alpha$  emission. It is possible that we are observing the so-called ``moving mirror'' effect, i.e., the dust nebula is receding with respect to the star, and its expansion velocity is added to the velocity of the H$\alpha$ line. A similar effect has been discussed, for example, for the nebula near T~Tau [19].

As for the HH~941 object, its knots have corresponding emission counterparts in the molecular hydrogen line [5]. These molecular condensations are also clearly visible in the 3.6 $\mu$m, 4.5 $\mu$m, and 5.8 $\mu$m images constructed from data obtained with the Spitzer Space Telescope as part of the SEIP project. There is indeed every reason to believe that together with the counterflow HH~942, which we did not observe spectroscopically, they form a single bipolar outflow [2,5]. All the more interesting, therefore, is the large difference in physical conditions and velocities between the two condensations. With regard to a possible increase in the absolute flow velocity with distance from the star, it is still too early to draw firm conclusions, although such an increase in outflow velocity is, in general, not uncommon.

With respect to the HH~940 object, one can only state that the probability of its association with CPM~19 is quite high. The absence of a visible in the optical range collimated jet, connecting them, is not a necessary condition for such conclusion. It should be noted, that in [5] numerous molecular hydrogen condensations detected around this source are described; among them, the objects MHO 734 A, D, and E are identified with the optical condensations of HH~940. In [5], a direct association of one of the molecular outflows with CPM~19 is also reasonably suggested (in total, three candidate molecular outflows around CPM~19 are identified in that work; see also their discussion in [6]). Unfortunately, it is not possible to compare these data directly with the SEIP images, since all these objects are lost in the halo of the extremely bright IR source CPM~19. 

Based on all of the above, despite the limited amount of observational material, we can draw the following conclusion: even if the visible and infrared sources located at the position of CPM~19 represent two different stars, the optical object is very likely also a YSO. This makes the object, which has been studied for decades, even more intriguing. Unfortunately, further progress in the investigation of CPM~19 is hampered by its faintness in the optical range. Apparently, significant advances in understanding the structure of this system and of this small star-forming region can only be achieved using the largest facilities, such as the VLA and ALMA.

\section*{Acknowledgements}
Authors thank the referee for very valuable comments and suggestions. 

Observations at the telescopes of the Special Astrophysical Observatory of the Russian Academy of Sciences (SAO RAS) are carried out with the support of the Ministry of Science and Higher Education of the Russian Federation. The upgrade of the instrumental base is performed within the framework of the national project \textit{Science and Universities}. The acquisition and reduction of the spectroscopic data were conducted as part of the state assignment of SAO RAS approved by the Ministry of Science and Higher Education of the Russian Federation. '

T.~Molyarova was supported by the Royal Society (URF\textbackslash R1\textbackslash211799, RF\textbackslash ERE\textbackslash231082).

The Pan-STARRS1 Surveys (PS1) and the PS1 public science archive were made possible through contributions by the Institute for Astronomy, the University of Hawaii, the PanSTARRS Project Office, the Max-Planck Society and its participating institutes, the Max Planck Institute for Astronomy, Heidelberg and the Max Planck Institute for Extraterrestrial Physics, Garching, Johns Hopkins University, Durham University, the University of Edinburgh, Queen's University Belfast, the Harvard-Smithsonian Center for Astrophysics, the Las Cumbres Observatory Global Telescope Network Incorporated, the National Central University of Taiwan, the Space Telescope Science Institute, NASA under grant no. NNX08AR22G issued through the Planetary Science Division of the NASA Science Mission Directorate, the National Science Foundation grant no. AST-1238877, the University of Maryland, Eotvos Lorand University (ELTE), the Los Alamos National Laboratory, and the Gordon and Betty Moore Foundation.

\end{document}